\documentstyle[twocolumn,prl,aps,floats]{revtex}

\begin{document}

\input{psfig.sty}

%\draft
%\wideabs{
\title{Solid-State Systems for the Electron Electric Dipole Moment
and other Fundamental Measurements}

\author{S.K. Lamoreaux}

\address{University of California,
Los Alamos National Laboratory, Physics Division, Los Alamos, NM 87545}

\date{\today}

\maketitle
\begin{abstract}

In 1968, F.L. Shapiro published the suggestion that one could
search for an electron EDM by applying a strong electric field
to a substance that has an unpaired electron spin; at low
temperature, the EDM interaction would lead to a net sample
magnetization that can be detected with a SQUID magnetometer.
One experimental EDM search based on this technique was published,
and for a number of reasons including high sample conductivity,
high operating temperature, and limited SQUID technology, the result
was not particularly sensitive compared to other experiments in
the late 1970's.

Advances in SQUID and conventional magnetometery had led
us to reconsider this type of experiment, which can be extended
to searches and tests other than EDMs (e.g., test of Lorentz invariance).
In addition, the complementary measurement of an EDM-induced
sample electric polarization due to application of a magnetic field
to a paramagnetic sample might be effective using modern
ultrasensitive charge measurement techniques.  A possible
paramagnetic material is Gd-substituted YIG which has very low
conductivity and a net enhancement (atomic enhancement times
crystal screening) of order unity.  Use of a reasonable
volume (100's of cc) sample of this material at 50 mK and 10 kV/cm
might yield an electron EDM sensitivity of $10^{-32}$ e cm or better,
a factor of $10^5$ improvement over current experimental limits.

\end{abstract}%} 
\pacs{11.30.Er, 77.22.-d,67.80.Jd}

\section{Introduction}

The idea to use  solid state systems for permanent
electric dipole moment (EDM) and
other fundamental measurements has been around for quite some
time; with the exception of one electron EDM measurement of rather
unremarkable sensitivity, (spin polarized) solid state systems have found
their only applications to Lorentz violation and new-long-range
force tests using torsion pendula; see  \cite{torsion} for an overview.

F.L. Shapiro, in 1968, put forward the idea that one could test
for the presence of 
an electron EDM by applying a strong electric field to a material
with unpaired electron spins;  the EDM of the sample atoms
(or ions), and therefore
the spins, become spin-polarized \cite{shapiro}. The degree of 
spin-polarization can
be determined, in conjunction with  the Boltzmann equation, 
from the hamiltonian
\begin{equation}\label{int}
H=-d{{\bf E\cdot J}\over J}
\end{equation}  
where $d$ is the EDM (measured in $e$cm), $E$ is the applied
electric field, and $J$ is the total atomic (or ionic) angular 
momentum.
Because each spin also carries a
magnetic moment, the sample will become magnetized.  The change in
magnetic flux $\Phi$ at the surface of a flat sheet of material 
with the application of an electric field is
\begin{equation}\label{phi}
\Delta\Phi=4\pi \chi A d E^* /\mu_a
\end{equation}
where $\chi$ is the magnetic susceptibility, $A$ is the sample
area, $d$ is the EDM associated with
the spin of interest, $E^*$ is the effective electric field at the location
of the spins of interest, and $\mu_a=g[J(J+1)]^{1/2}\mu_B$ is atomic or
ionic  magnetic moment.  The diamagnetic susceptibility
is determined by the ratio of the magnetization $M$ to the
magnetic induction $B$
\begin{equation}\label{B}
\chi={M\over B}\approx {N\mu_a^2\over 3 k_BT}
\end{equation}
where $N$ is the number density of spins of interest,
$k_B$ is Boltzmann's constant, and $T$ is the sample temperature,
in the cases where simple Langevin paramagnetism is applicable
(which we will assume for demagnetized soft ferro/ferrimagnetic
materials).

We should note that any hamiltonian of the form Eq. (\ref{int}) 
will lead to a sample magnetization, e.g., a Lorentz violation
would lead to a sample orientation dependent magnetization, or
some new long range force would lead to a magnetization that 
depends on the separation between the paramagnetic material
and a laboratory source.  Such experiments will not be
directly discussed here; the possible improvements
for limits on such interactions
can easily be determined by scaling the potential electron EDM limits
presented here with previous EDM results.  

\section{Previous Work}

The one experiment to measure an electron EDM by the Shapiro
technique employed a nickel-zinc ferrite which could support
an electric field of 2kV/cm and had high resistivity at
low temperature \cite{exp}.  The ion of interest in this case is
Fe$^{3+}$.  Because of the low atomic number of Fe, the
net EDM of the atom is $d=0.5 d_e$, where $d_e$ is the
bare electron EDM.  

One possible source of shielding
of the electric field in the crystal was neglected
in this work.  As is well-known in the case of
atoms, the net electric field at the site of the nucleus
and at each electron in the atom must be zero for the
system to be in equilibrium (when only electrostatic
forces are present).  In the case of a bulk crystal,
 $E^*$ is determined by the
average of $\langle {\bf d\cdot E}\rangle$ over the position of
the ion in the crystal lattice.  We must recall that
for the system to be in equilibrium, when only
electrostatic forces are present, the average of
$\langle q(r)E(r)\rangle$, e.g., the atomic charge density 
times the electric field (which varies rapidly in the crystal),
must be zero.  This effect was not taken into account
in \cite{exp}, but is likely not a large effect.
This is because a system bound only by electrostatic
forces is not stable, so there are additional
forces within the crystal.  Exchange forces (which represent
dynamic quantum fluctuations in the fields) that ultimately keeps
the ions separated are
roughly as important as the electrostatic interactions
between ions in a crystal.  This is equivalent
to saying that if an ion is ``large'' it really can't move
much in the crystal, and the displacement due to the
applied electric field is against the exchange force
which has a different spatial functional dependence
compared to the crystal electrostatic forces.

However, there is
an effective  screening due to the high dielectric
constant which was taken into account
in \cite{exp}.  The experimental result, $d_e=-8.1\pm 11.6\times 10^{-23}$ 
represent a magnetic
sensitivity of $3\times 10^{-11}$ G which was obtained in 3.5
hours with a SQUID magnetometer, with the experiment operated
at 4 K. The flux detection sensitivity corresponds to about 5 $\mu \Phi_0$
where $\Phi_0=2.07 \times 10^{-7}$ G-cm$^2$. 
We can
conclude that the SQUID sensitivity was 800 $\mu \Phi_0/\sqrt{\rm Hz}$.
In addition,
$A$ in Eq. (\ref{phi}) is coupled to a SQUID that has a small area,
resulting in an effective sampling area $2\times 10^{-2}$ smaller
than the ``true'' sample area.
As will be discussed, this is a general problem with SQUID
magnetometery.

Further experiments with EuS and EuO proved unsuccessful due to
large leakage currents with application of the high voltage.

\section{A Modern Solid State SQUID EDM Experiment}
\subsection{Materials}

For a modern experiment, we are considering garnet crystals.
The iron and gadolinium rare earth garnets are cubic insulating
crystals so they have sufficient symmetry to suppress quadratic
effects \cite{dzy}.
The rare earth iron garnets (RIG, where R is Y, Gd, etc.,
and is in the R$^{3+}$ ionization state)
are ferrimagnetic while
the rare earth gallium garnets (RGG) are Langevin paramagnets (Ga has
no paramagnetism)
and follow the Brillouin formula Eq. (\ref{B}) \cite{path}.  For the
case of an EDM induced magnetization, the ferrimagnetism 
presents some complications, so for the arguments presented
in this section, GdGG will be considered.
The formula of this material is Gd$_3$Ga$_5$O$_{12}$, which represents
the general garnet formula  where
other rare earth elements, either singly or in mixtures,
can be in place of some or all of the Gd;
Ga is replaced by Fe for RIG.  All these materials have volume resistivities
of greater than $10^{16}\ \Omega$cm at temperatures at or below
77 K.  The issues of dielectric strength remain to be investigated.
Also questions of whether spin-glass or ferromagnetic transitions
occur remain to be investigated (note that GdGG is used in
adiabatic demagnetization refrigerators).  

Gd$^{3+}$ is experimentally attractive because it has total angular momentum
$L=0$ and total spin $S=7/2$.  The lack of orbital
angular momentum makes it easy to magnetize
the material.  Ga has no magnetic moment so
Gd$^{3+}$ determines all the magnetic properties
of GdGG. The electronic configuration in Gd$^{3+}$ is complicated,
but it has at least one $6s$ valence electron;  we might expect 
a net atomic enhancement of $d=\alpha^2 Z^3\approx 10$, while
the shielding of the electric field in the crystal might be around an
order of magnitude.  A preliminary estimate by O. Sushkov and S. Kuenzi
\cite{sush} for GdIG shows a net enhancement times screening of order unity.
We can thus assume, in Eq. (\ref{phi}) that $dE^*=d_eE$ where
$E$ is the electric field applied to the crystal.

The density of Gd in GdGG is about 10$^{22}$/cc, somewhat low,
but given the excellent insulating properties, this is
an acceptable sacrifice.  From Eq. (\ref{B}), the 
assuming an electron EDM of $10^{-27}$ e cm and an electric field of
10 kV/cm implies
\begin{equation}
\Delta \Phi=d_e
{3.16 \times 10^{-16}\over T} {GA\over 10^{-27}{\ \rm e\ cm}\ \  10
{\rm\ kV/cm}}
\end{equation}
so, for $T=10$ mK, $A=100\ {\rm cm^2}$, $\Delta\Phi=17\ \mu\Phi_0/d_e$.

\subsection{Magnetometry}

The best-possible energy resolution that a SQUID magnetometer
can achieve is dictated by the energy uncertainty principle,
$\Delta E\Delta t\geq \hbar$.  A  useful way to parameterize
the sensitivity of a SQUID is by its intrinsic energy resolution
$d E_{sq} t = n\hbar$, where $n\geq 1$,
relative
to a perfect SQUID;
\begin{equation}\label{ssens}
d E_{sq}t={d\Phi_{sq}^2\over 2 c^2 L_{sq}} =n\hbar
\end{equation}
where $d\Phi_{sq}$ is the flux through the SQUID loop,
$L_{sq}$ is the intrinsic inductance of the SQUID, and $t$
is the measurement time.
Modern SQUID magnetometers routinely achieve an intrinsic energy 
sensitivity of
order $10\hbar$.
For comparison, the SQUID used with the work described in
\cite{exp} was
of order $6.4\times 10^7\hbar$.  

Equation (\ref{ssens}) 
represents the energy noise per second integration time $t$,
implying a flux sensitivity of
\begin{equation}
d\Phi={\sqrt {2n\hbar c^2 L_{sq}/t}}\approx 0.2 \mu\Phi_0\sqrt{{\rm\ sec/t}}
\end{equation}
where $t$ is the integration time.
For modern commercially-available SQUIDs, 
typically $L_{sq}=
10^{-12}\ \rm{s^2/cm}=0.2$ nH, 
with an input coupler inductance of
$L_i=500$ nH, and the input  mutual inductance is $M=\sqrt {L_{sq}L_i}=10$
nH. (For convenience, we
will use MKSA inductance
units for describing
SQUID properties.) 
Thus, the fraction of flux picked-up from the sample that is
delivered to the SQUID is
\begin{equation}
d\Phi_{sq}=d\Phi_p {M\over L_p +L_i}
\end{equation}
where $L_p$ is the inductance of the pickup coil around the sample.
Therefore we see a loss of sensitivity over that intrinsic to the SQUID;
this is due to the area mismatch between the sample and SQUID areas.
We could imagine building a SQUID with a lower input inductance, 
perhaps achieving $L_i=L_{sq}=M$.  In this case, $L_i$ is
much smaller than any imaginable $L_p$; because $L_p$ scales
as the diameter of the sample,
\begin{equation}\label{induct}
L=6.27\times 10^{-3} D\left[\ln {D\over d}-2\right]\ \ \mu{\rm H}
\end{equation}
where $D$ is the diameter of a circular pickup loop in cm, and $d$ is
the diameter in cm of the superconducting 
wire used in its construction \cite{refdata}.
Because $\Delta \Phi_p$ scales as the sample area (proportional to
$D^2$),
while the inductance scales as the sample diameter $D$, we see
the sensitivity scale linear in the sample size, or sample
volume $V^{1/3}$; in the case that $L_p<<L_i$, the scaling
is $V^{2/3}$.  

With a $10\hbar$ SQUID magnetometer, with a 100 nH pickup coil around a
100 cm$^2$ area sample, we might expect a sensitivity of
\begin{equation}
d_e\approx 
(0.2\mu\Phi_0
\sqrt{\rm \ sec})
\left[{17\mu\Phi_0\over 10^{-27}\ e{\rm cm}}{M\over L_i+L_p}\right]^{-1}
\end{equation}
\[
=0.7\times 10^{-27}\ e{\rm cm}\sqrt{\rm\ sec}
\]
which leads to a sensitivity of $10^{-30}$ e cm in 10 days
of averaging.

However, it is possible to do much better.  Magnetometry based on
the non-linear Faraday effect in atomic vapors \cite{dima}  has produced a
sensitivity of $3\times 10^{-12} \rm{G/\sqrt{Hz}}$ and might
be improved by several orders of magnitude by producing and
interrogating atoms contained with a cold dense buffer
gas \cite{dima2}.  The improvement comes
from a narrowing of the magnetic
resonance lines to a few milli-Hz.  This system
has the advantage that the large sample can be conveniently 
matched to the magnetometer; if we assume the magnetometer volume
diameter is 1 cm, and the sample diameter is 10 cm, a superconducting
transformer can be used to pick-up the sample
magnetization and then step-up the magnetic
induction at the magnetometer.
From Eq. (\ref{induct}), the ratio of the inductances of
the two coils is roughly 1/10; the magnetic induction at the
center of the
small coil is 10 times that of the large coil, with a given
current in the series-connected coils.  The magnetometer would
have to be operated near a temperature of 2 K, while the
sample is at 0.01 K.  This would pose no problem because
the superconducting transformer connection can be between two
regions that are at different temperatures; the thermal
conductivity of superconductors is very small so the heat load
can be controlled.

With the factor of 10 due to magnetic induction step-up, along with
a two-order-of-magnitude increase in sensitivity due to
the line width improvement, 
we could expect a magnetic induction sensitivity of
$3\times 10^{-15}\ {\rm G/\sqrt{Hz}}$,  or an EDM sensitivity
of $10^{-29}\ \rm{e cm/\sqrt{Hz}}$; in ten days of averaging,
the sensitivity is $10^{-32}$ e cm.

As a comparison, a direct measure of the EDM of the magnetometer
atoms, by applying 10 kV/cm, would be around $5\times 10^{-31}$ e cm,
a factor of 200 worse than one can obtain by measuring the
induced magnetization of the solid system.

By operating the experiment at an even lower temperature,
say 10 $\mu$K which is not technically impossible, a
sensitivity of $10^{-35}$ e cm is conceivable.

\subsection{Systematics}

The usual systematics that one encounters with EDMs based on magnetometry
with atomic vapors will be present for the solid state experiment.
The leakage current danger in this case is a magnetic field that
is picked up directly by the SQUID or Faraday magnetometer.  As an example,
a 10 cm diameter quarter-turn  leakage current of $10^{-14}$ A, which might
be expected at low temperatures with 10 kV/cm and a large
sample,  corresponds to
a spurious field of $1\times 10^{-15}$ G, or 2,000 times the
expected 10-day sensitivity.  Even more worrisome is the displacement
current magnetic field; if the electric field is reversed at 10 Hz (to avoid
the $1/f$ corner of SQUID magnetometers) the displacement current is
10 $\mu$A for a 100 pF sample, assuming $E=10$ kV/cm.  
The sample magnetization would be measured after
the high voltage has stabilized, but the displacement current
magnetic field is so enormous that we can be concerned about
hysteretic or other non-linear effect.  Another limitation
to the reversal frequency is the spin-lattice relaxation time--
this can be measured using standard techniques.  Another concern is
energy dissipation in the sample with electric
field reversal, which, when the experiment
is performed at 10 mK, must be limited to 10 $\mu$W as set by
the cooling power of a typical dilution refrigerator at low temperature.
The $1/f$ corner of a Faraday magnetometer might occur at much
lower frequency,
allowing less frequent field reversals.  The high voltage
properties of materials remain to be studied; the number presented
here show some limits to the technique.  Clearly if one applied
100 V to the sample, an EDM sensitivity of $10^{-30}$ e cm could
be achieved, and the leakage and displacement current problems
would be reduced to a manageable level. 

A separate class of systematics arise from macroscopic parity
and time-reversal odd effects due to the crystalline
structure.  Such effects have been predicted \cite{dzy} and
observed \cite{astro1,astro2} in non-centrosymmetric single
crystals.  These effects are absent for symmetric crystals
but a realistic system will always have strains and imperfections.
Use of a polycrystalline sintered sample would tend to randomize
these effects.

\section{Magnetization-Induced Sample Electric Polarization due to 
an EDM}

\subsection{Introduction}

As suggested by D. DeMille \cite{dave}
if the paramagnetic atoms responsible for the magnetic properties
of a material also have an EDM, then with the sample is
magnetized, there will be an induced sample electric polarization.
If the sample is ferromagnetic, the coercive 
magnetic field $H_c$ that must
be applied to attain the remnant saturation magnetization is
fairly small, 20-500 Oe.  Materials such as
GdIG or GdYIG might be of considerable interest; more will
be said about their properties later.

Let us estimate the size of the induced electric polarization.  Again,
take the present electron EDM limit of $10^{-27}$ e cm as
the characteristic EDM scale.  Also, take the density of
Gd in GdIG as $\rho=10^{22}$/cc. The induced electric field
is calculated as
\begin{equation}
{\cal E} =4\pi\rho d P=6\times 10^{-14}{\rm \ statvolt/cm}
\end{equation}
\[=1.8\times 10^{-11}
{\rm \ V/cm}.
\]
where $P$ represents the degree that the spins are polarized in
the sample, and it is possible $P\approx 1$ for ferro- and ferri-magnetic
materials.
The voltage across a capacitor is obtained by
multiplying $\cal E$ by the sample length $L$ and dividing by
by the dielectric constant $\varepsilon$. 
For GdIG etc., $\varepsilon\approx 15$; 
we might also expect that $d\approx 15 d_e$ as discussed
previously. 
We can  assume that the atomic enhancement roughly
cancels the reduction in voltage due to the dielectric constant.
Therefore, if we consider a sample 10 cm thick, the EDM induced
voltage for an electron EDM of
$10^{-27}$ e cm  will be 0.18 nV, assuming $P=1$.

\subsection{Voltage Measurement}

We are interested in measuring sub-nanovolt voltages at
modulation frequencies in the range 10-100 Hz, a practical
range considering the time to reverse the sample magnetization
in a controlled way, and allow for spurious electric field
associated with the time varying magnetic fields to dissipate.

For reasons outlined below, a total sample capacitance of about
100 pF would be ideal for this experiment; this sample would be
10 cm thick and 30 cm diameter, and likely represents a practical
maximum for sample preparation.  

The parallel input resistance of the sample and amplifier, together
with the summed capacitances, should give an input time constant
much longer that the inverse modulation frequency; 
\begin{equation}
\tau_c=(R_s||R_a)(C_s+C_a)
\end{equation}
where $R_{s,a},C_{s,a}$ are the sample and
amplifier resistances and capacitances, respectively.
The Johnson (voltage) noise on the amplifier input is, ignoring
the total capacitance
\begin{equation}
V_n(0)=(R_s||R_a)\sqrt{{4k_BT_s\over R_s}+ {4k_BT_a\over R_a}}
\end{equation}
where the possibility for the amplifier and sample
temperatures to be different has been explicitly
included; a practical limit for
the amplifier is 120 K (see below).  The net time constant
at the amplifier input will limit the frequency response;
\begin{equation}
V_n(\omega)={V_n(0)\over \sqrt{1+(\omega\tau_c)^2}}\approx {V_n(0)\over
\omega\tau_c}
\end{equation}
where we assumed $\omega\tau_c>>1$. The noise
is, assuming  $10^{15}\Omega$ for net amplifier and sample resistance
(with the amplifier and sample at the same temperature), is
$V_n(0)=2 {\rm mV} {/\sqrt{\rm Hz}}$; if the modulation frequency is
10 Hz, with $\tau_c=10^{15}\Omega\times 100 {\rm\ pF} =10^5$ sec,
we have $2{\rm mV/\sqrt{Hz}} /(2\pi f \tau_c)=0.3 {\rm nV/\sqrt{Hz}}$.

The combined requirement of high amplifier input
resistance, low $1/f$ noise, and low bias current imply the use
of a JFET input amplifier.  The $1/f$ noise, input resistance, and
input bias current are all a consequence of the same physical mechanism:
the generation current from Shockley-Read-Hall  generation-recombination (g-r)
centers in the gate {\it p-n} junction depletion region \cite{lor}.  The $1/f$ corner
is determined by the g-r time constant, which at room temperature for
modern JFETs is of order 1 msec, and varies with
temperature as
\begin{equation}\label{gr}
\tau_{gr}=\tau_0e^{E/k_BT}.
\end{equation}
The total noise voltage is determined by
the square root of the total number of g-r centers (impurities) in the 
depletion region; because the total number depends on the depletion
region volume and the depletion depth is constant independent of
specific JFET, the $1/f$ (and some other types of noise) 
noise voltage scales inversely as the square root of the
depletion region area (roughly the gate area, hence gate
capacitance).  The g-r center density
is a function of preparation technology and therefore also independent
of specific JFET, for modern low-noise devices.  
This scaling is accurate as can be seen by
comparing the intrinsic $1/f$ noise of various JFETs as a function
of gate capacitances.  

JFETs can be operated to temperatures down to about $T_{min}=$120-140 K; $g_m$,
the transconductance, continues to increase until that temperature,
and for lower temperatures, the carriers freeze out so $g_m$ 
decreases, and the noise increases.
There is a modest decrease in device noise with decreasing
temperature due to the $\sqrt T$ dependence
of Johnson noise; however, the g-r time constant increases exponentially.
For many devices, the noise becomes frequency independent (above 10 Hz)
at temperatures
around $T_{min}$ and is about a factor of two lower than the room
temperature high frequency noise.  The JFET gate capacitances are
almost independent of temperature.

If we choose a sample capacitance, the amplifier capacitance to optimize
the signal-to-noise (assuming the noise is only due to
the JFET) can be determined by minimizing
\begin{equation}
{S\over N}\propto{(C_a+C_s)^{-1}\over (C_a)^{-1/2}}
\end{equation}
which has a maximum when $C_a=C_s$.  If $C_a$ is fixed, the optimum sample
capacitance can be determined; the sample capacitance
is $C_s=\epsilon A/L$, $A=$ area, $L=$length.  Therefore the optimum $S/N$ is
(the total charge on each end of the polarized 
sample is proportional to $\pm dA$,
independent of $L$)
\begin{equation}
S/N\propto{{A( C_a+\epsilon A/L)^{-1}}\over {{(C_a)^{-1}}}}\propto 
L\sqrt{C_a}/2
\end{equation}
when $C_a=C_s$, implying $L$ and $C_a$ should be as large as possible.
The dimensions given above for a 100 pF sample represent a practical
maximum.  After choosing a sample, $C_a$ can be set by choice
of JFET, or by placing several JFETs in parallel (the capacitance
increases linear with the number in parallel $N_j$, while the noise
decreases as the square root $\sqrt{N_j}$, which in consistent
with the early discussion of JFET noise).  The overall sensitivity
scales as $L\sqrt{A}\propto V^{2/3}$.

A JFET that might be useful in this application is the Interfet IF3601
(or the IF3602 dual JFET model).  This device has a rather high input
capacitance $C_{iss}=$300 pF and and high reverse
transfer capacitance $C_{rss}=$200 pF, but with noise
of 0.3nV$/\sqrt{\rm Hz}$ at 100 Hz.  In a properly design cascode
configuration \cite{oxner}, the 
input capacitance will be $C_{iss}-C_{rss}=$100 pF,
with the intrinsic noise unmodified \cite{walls}.  If this device
is cooled to 120 K, we might expect a noise of 0.2nV$/\sqrt{\rm Hz}$
at 10 Hz and above.  Thus, the EDM sensitivity is (neglecting a possible
0.3nV$/\sqrt{\bf Hz}$ Johnson noise mentioned above; this can
be reduced by operating at higher than 10 Hz, or if the
net input resistance is $10^{16}\Omega$ or larger)
$d_e=10^{-27}{\rm e\ cm}/\sqrt{\rm Hz}$, or a sensitivity
of $10^{-30}$ e cm in ten days of operation.

\subsection{Materials}

GdIG might be an ideal material for this type of experiment.  At 4 K,
the Gd ion spin is completely polarized by the ferrimagnetic sublattices.

It should be noted that the magnetic susceptibility of GdIG is
very temperature dependent, and is zero at a specific
temperature near room temperature.
The point where the magnetic susceptibility is zero is called
the compensation temperature $T_c$ and results from the polarization
of the Gd ions in the lattice exactly canceling the magnetization
of the Fe$^{+2}$ and Fe$^{+3}$ paramagnetic ions (see, e.g.,
\cite{kittel} for an excellent discussion).  By mixing in Ytterbium
which carries no magnetization, the compensation temperature
can be adjusted.  With approximately 1:2 Gd:Yb, $T_c\approx 77$ K.
An  advantage to use of such a material is that the
sample magnetization is very small, reducing some possible systematic
effects \cite{hunter1}.  On the other hand the
coercive field is roughly given
by $H_c=  250 /(|T_c/T-1|+1)$ Oe for polycrystalline GdYIG and the
increase near $T_c$ is due to 
the low $B$ for a given $H$ when the 
magnetization of the material is
small; this describes the 
primary dependence of $H_c$ on $T$ and the composition
 \cite{hanton}.   Another interesting
point is that the hysteresis
loop becomes very square near $T_c$; this is attributed to 
the domains being very large when the magnetization of the
sample is small \cite{nelson,sano}. 
The switching speed is determined by the domain wall velocity,
which is typically $10^4$ cm/s.  Thus, a sample of 10 cm length
could be switched in 1 ms.  

The Yb concentration is determined by the operating temperature,
which is chosen so that the sample resistivity is high.  A convenient
temperature is $T=77$ K, corresponding to liquid nitrogen at 1 bar pressure;
as discussed before, the Gd density is reduced by a factor of three
to achieve $T_c=77$ K.  Also, the Gd polarization as given
by the Boltzmann distribution at 77 K is reduced by
a factor of 2/3 compared to $T<4$ K.  Thus, the signal is reduced by
a factor of nearly 5 compared to an
``ideal'' experimental situation.  
(Note that we must also have $T\neq T_c$ so that
there is some magnetization; $T<T_c$ by a few
K should be sufficient.)

\subsection{Systematic Effects}

In order to reverse the magnetization of a sample, a magnetic pulse with
amplitude larger than $H_c$ must be applied to the sample for a time
somewhat longer than the switching time estimated above as 1 ms,
after which the field can be reduced to a value just sufficient
to ``hold'' the magnetization. The
switching rate is also determined by the $1/f$ corner of the amplifier;
the transient effects favor a lower rate of magnetization
reversal ($< 1$ KHz), while the $1/f$ corner favor a faster rate
($>100$ Hz).  Thus, a reversal rate of a few hundred Hz satisfies
both constraints.

There are induced voltages associated with the reversal pulse,
and the system must be carefully designed to avoid overloading
the amplifiers. In addition, slowly decaying eddy currents etc.
could create a time-dependent induced voltage that masks or mimic an
EDM.  The time-variation of the EDM  signal should be an exponential
decay with time constant $\tau_c$ given above.  
If the switching spurious signal is rectified by the amplifier
input, the effect would look exactly like an EDM signal.

Another concern is magnetostriction; if there is a permanent
magnetized strain in the material, there could be a magnetization
dependent distortion of the sample.  One fixed electron coupled with
a change in sample dimension of $10^{-4}$ cm would give a $10^{-30}$ e cm
signal.  
 
\section{Conclusions}

We have described a series of solid state based EDM experiments that
offer over a factor of 1000 improvement on the electron EDM
limit, and perhaps up to a factor of $10^6$ improvement
when a new type of magnetometer
is employed.  These experiments are intrinsically ``easy'' compared to
the more traditional atomic cell or beam resonance experiments, and
suffer from many of the same systematics.  Given that the
experiments proposed here are not terribly elaborate, they would be worth
doing simply to see what happens.  There are a number of
issue to be studied; these include the following:  atomic (ionic)
enhancement of $d_e$ in Gd and other rare earths; screening of
the applied electric field at the ion locations; screening of
the induced dipole in the voltage measurement experiment;
properties of materials, in particular dielectric strength and
leakage currents as a function of applied voltage; and resistivity
as a function of temperature.  The various garnet materials
might not be the best choice due to the relatively low rare earth density.
Perhaps some other materials would be a better choice.  The garnets
were chosen here for discussion because it is well known that they
have excellent insulating properties.

As an interesting aside, an experiment using a rotating
YIG rod to test special relativity
was performed at Amherst College \cite{larry}; in this experiment,
a voltage was induced between the ends of the rod due to the
rotation.  Agreement with special relativity was obtained, and
required measurement of voltages at the mV level with $\mu$V accuracy.  Given
the success of this experiment, the work proposed here does not
seem unreasonable.  

The proposed techniques also can be applied to other fundamental
studies.  For example, if the nucleus has an anapole moment,
the electron cloud around the nucleus will be displaced.
This is because 
the low-energy parity violating weak interaction, being
proportional to the momentum $\hat p$ can be
written as the operator of parallel displacement
(of the electron cloud relavite to the nucleus)
with an imaginary amplitude \cite{landau}.
If the
nuclear spins of a material are polarized
in a strong magnetic field at low temperature, and if
the direction of the nuclear spins is reversed with an NMR
pulse, there will be a change in voltage across the sample, as measured
along the magnetic field.  In this case, the parity-odd
time-reversal-even operator describes the displacement of
charge along a specified that is correlated with the sample polarization
which  is established by an applied magnetic field,
$\hat p\cdot \hat B\propto
 i\hat r \cdot \hat B$; the factor of $i$ changes sign under
time-reversal.  Hence, the operator is $T-$even and $P-$odd.
The magnitude of this voltage
can be estimated as a few $\mu$V using expected values of
nuclear anapole moments.

In conclusion, there is hope to improve the electron
EDM experimental limit by at least three orders of magnitude by
use of the experimental techniques proposed here. 
By operating the system proposed in Sec. III at a temperature
of 10 $\mu$K, a sensitivity of $10^{-35}$ e cm is not beyond
conception.  Such a sensitivity is several orders of magnitude
better than any currently proposed experiment.


\begin{thebibliography}{99}

\bibitem{torsion} R. Bluhm and V.A. Kostelecky, Phys.
Rev. Lett. {\bf 84}, 1381 (2000).

\bibitem{shapiro} F.L. Shapiro, Sov. Phys. Usp. {\bf 11}, 345 (1968).

\bibitem{exp} B.V. Vasil'ev and E.V. Kolycheva, Sov. Phys. JETP
{\bf 47}, 243 (1978).

\bibitem{dzy} I. Dzyaloshinskii, Sol. State Comm. {\bf 82}, 579 (1992).


\bibitem{path} L. N\'eel, R. Pauthenet, and B. Dreyfus,
Prog. Low Temp. Phys. {\bf 4}, 344 (1964).  See Fig. 9.

\bibitem{sush} O. Sushkov, Private Communication, 2001;
S. Kuenzi and O. Sushkov, to be published, Phys. Rev. A.

\bibitem{refdata} F.E. Terman, {\it Reference Data for Radio
Engineers} (McGraw-Hill, New York, 1943).

\bibitem{dima} D. Budker et al., Phys. Rev. A {\bf 62}, 043403 (2000).

\bibitem{dima2} D. Budker, Private Communication, 2001.


\bibitem{astro1} D.N. Astrov and N.B. Ermakov, JETP Lett.
{\bf 59}, 297 (1994). (Eng. Trans.) 

\bibitem{astro2} D.N. Astrov et al., JETP Lett. {\bf 63}, 745 (1996).
(Eng. Trans.)

\bibitem{dave} D.P. DeMille, Private Communication, 2001.

\bibitem{lor} P.O Lauritzen, Solid State Elec. {\bf 8}, 41 (1965).
C.T.Sah, IEEE Trans. on Electron Devices {\bf ED-11}, 795 (1964).

\bibitem{oxner} E.A. Oxner, {\it Designing with Field-Effect
Transistors, 2nd Ed.}, pp. 89-93 (McGraw-Hill, New York, 1990). 


\bibitem{walls} Steven R. Jefferts and F.L. Walls, Rev. Sci. Instr.
{\bf 60}, 1194 (1989).

\bibitem{kittel} Charles Kittel, {\it Introduction to Solid State
Physics, 4th ed.} (Wiley, New York, 1971).  See Chap. 16.

\bibitem{hunter1} L.R. Hunter, Private Communication, 2001.

\bibitem{hanton} John P. Hanton, IEEE Trans. on Magnetics {\bf MAG-3},
505 (1967).

\bibitem{nelson} Thomas J. Nelson, IEEE Trans. on Magnetics {\bf MAG-4},
421 (1968).

\bibitem{sano} Antonio San\`o and Antonio Serra, IEEE Trans. on Magnetics
{\bf MAG-4}, 646 (1968).

\bibitem{larry} J.B. Hertzberg et al., Am. Jour. Phys. {\bf 69}, 648 (2001).

\bibitem{landau} L.D. Landau and E.M. Lifshitz, {\it Quantum
Mechanics, Third Edition}, p. 45 (Pergamon, New York, 1977).
 
\end{thebibliography}
\end{document}